\documentclass[11pt]{article}
\usepackage{fleqn,cospar}
\usepackage{times}
\usepackage{cospar}
\usepackage[sectionbib]{natbib}
\renewcommand{\bibsection}{\section*{REFERENCES}}
\renewcommand{\bibhang}{\setlength{8mm}}
\renewcommand{\bibsep}{\setlength{0mm}}
\usepackage{url}

\usepackage{graphicx}
\usepackage[figuresright]{rotating}

\title{CHANDRA OBSERVATIONS OF SNR 1987A}

\author{S. Park\address{Dept. of Astronomy and Astrophys, 
        525 Davey Lab., Penn State Univ., University Park, PA 16802, USA},
        S. A. Zhekov\address{Space Research Institute, Moskovska str. 6,
	Sofia - 1000, Bulgaria},
	D. N. Burrows$^{1}$,
	E. Michael\address{JILA, Univ. of Colorado, Campus Box 440, Boulder, 
        CO 80309, USA},
	R. McCray$^{3}$, G. P. Garmire$^{1}$,
        and
        G. Hasinger\address{Max-Planck-Institut f\"ur Extraterrestrische 
	Physik, Postfach 1312, D-85748, Garching, Germany}} 

\begin{document}
% typeset front matter
\maketitle
\begin{abstract}
We report on the results of our monitoring program of the X-ray 
remnant of supernova 1987A with the {\it Chandra X-Ray Observatory}. 
We have performed two new observations during the {\it Chandra} 
Cycle 3 period, bringing the total to six monitoring observations over 
the past three years. These six observations provide a detailed 
time history of the birth of a new supernova remnant in X-rays. 
The high angular resolution images indicate that soft X-ray bright 
knots are associated with the optical spots, while hard X-ray
features are better correlated with radio images. We interpret this
in terms of a model in which fast shocks propagating through the 
circumstellar HII region produce the hard X-ray and radio emission, 
while the soft X-ray and optical emission arise in slower shocks 
entering into dense knots in the circumstellar inner ring. 
New observations begin to show changes in the morphology that may 
herald a new stage in the development of this incipient supernova 
remnant. The observed X-ray fluxes increase by nearly a factor of
three over the last 30 months. The X-ray remnant is expanding at a 
velocity of $\sim$5000 km s$^{-1}$.

\end{abstract}
%\section*{TITLE, AUTHORSHIP, AND ABSTRACT PLACEMENT}
%\section*{INSTRUCTIONS FOR MANUSCRIPT TEXT}
\section*{INTRODUCTION}

With a known age, distance, and the progenitor type, supernova (SN) 
1987A provides an unprecedented opportunity for the study of the early 
evolution of a supernova remnant (SNR). The optical inner ring, as 
observed by the {\it Hubble Space Telescope (HST)} is believed to be the 
dense circumstellar medium (CSM) produced by the stellar winds from 
the massive progenitor (Luo and McCray, 1991). Inside of this inner ring 
is an HII region generated by the UV radiation from the progenitor 
(Chevalier and Dwarkadas, 1995). The development of optically bright 
spots along the inner ring (Pun et al., 1997; Garnavich et al., 1997) 
was interpreted as emission from the radiative shock where the 
blast wave begins to strike inward protrusions of the dense inner ring 
(Michael et al., 2000). In X-rays, SN 1987A was detected with the R\"ontgen 
Satellite ({\it ROSAT}) as an unresolved source and the observed X-ray 
flux was linearly increasing (Hasinger et al., 1996). The high angular 
resolution images from the {\it Chandra} observations have revealed 
a shell-like structure of the X-ray remnant (Burrows et al. 2000). 
Monitoring observations with {\it Chandra} have shown a
continuous development of the X-ray bright spots, which are in general
spatially coincident with the optical and radio spots; a steady increase 
of the soft X-ray flux; and evidence of the radial expansion
of the SNR (Park et al., 2002). The thermal X-ray spectrum was fitted
with an electron temperature of $kT$ $\sim$ 2.5 keV. The broadening
of the atomic emission lines as detected with the dispersed spectrum 
indicated a shock velocity of $\sim$3400 km s$^{-1}$, which provided 
direct evidence for electron-ion non-equilibrium behind the shock 
front (Michael et al., 2002). We have recently performed two new 
observations with {\it Chandra}, which bring the total of monitoring
observations of SNR 1987A up to six. We here report an update on the
results from all six monitoring {\it Chandra} observations of SNR 1987A.
\clearpage

\begin{table}[h]
\vspace{-5mm}
\begin{minipage}{75mm}
\section*{OBSERVATIONS AND DATA REDUCTION}

As of May 2002, we have performed a total of six monitoring observations
of SNR 1987A with the Advanced CCD Imaging Spectrometer (ACIS) on board
the {\it Chandra X-ray Observatory} (Table~1). The description of the
observations and the data reduction of the first four observations can
be found in literature (Burrows et al., 2000; Park et al., 2002). We 
have performed two new observations on December 12, 2001 and May 15, 
2002 during {\it Chandra} Cycle 3. The ACIS-S3 was used without gratings. 
We have processed these two observations in the same way as four previous 
observations: i.e., we have corrected the charge transfer inefficiency 
(CTI; Townsley et al., 2000) with the methods developed by Townsley et al. 
(2002) before further standard data screenings, the sub-pixel resolution 
method (Tsunemi et al., 2001) was applied, and then the image was
deconvolved (Burrows et al., 2000; Park et al., 2002 and the references 
therein) to achieve the best-utilizable angular resolution.
\end{minipage}
\hfil\hspace{\fill}
\begin{minipage}{95mm}
\vspace{-8mm}
  \caption{ ~List of the Chandra Observations of SNR 1987A}
\vspace{3mm}
\begin{tabular}{|ccccc|}
\hline
ObsID & Date & Instrument & Exposure & Src \\
         & (Age$^{a}$)      &           & (ks) & Cts\\
\hline
124+1387$^{b}$ & 1999-10-06 & ACIS-S3 & 116 & 690 \\
& (4609) & +HETG & & \\
122 & 2000-01-17 & ACIS-S3 & 9 & 607 \\
    & (4711) & & & \\
1967 & 2000-12-07 & ACIS-S3 & 99 & 9031 \\
    & (5038) & & & \\
1044 & 2001-04-25 & ACIS-S3 & 18 & 1800 \\
    & (5176) & & & \\
2831 & 2001-12-12 & ACIS-S3 & 49 & 6226 \\
    & (5407) & & & \\
2832 & 2002-05-15 & ACIS-S3 & 44 & 6429 \\
    & (5561) & & & \\
\hline
\end{tabular}\\

$^{a}$ Day since the SN explosion in the parentheses\\
$^{b}$ The first observation was split into two sequences, which
were combined in the analysis.
\end{minipage}
\end{table}
\section*{X-RAY IMAGES}

The broad-band {\it Chandra}/ACIS images from six monitoring observations 
of SNR 1987A are presented in Figure~1. The overlay contours are from {\it
HST} H$\alpha$ data. The observation dates of the {\it HST} data are presented 
in the parentheses. The X-ray remnant is shell-like, as previously reported 
(Burrows et al., 2000; Park et al., 2002). Although the asymmetric surface 
brightness between the east and the west is persistent for all six images, 
the developements of new X-ray spots in the western half are evident in 
the latest images. These developing X-ray spots are spatially correlated 
with the optical spots. These ``simultaneous'' developments of the X-ray 
and optical spots support that the X-ray bright spots are where the blast 
wave is decelerated as it approaches the dense protrusions of the inner 
ring. We may thus expect the emergence of a {\it complete} X-ray ring 
in the near future as the shock front eventually strikes the main body of 
the entire inner ring. Park et al. (2002) have reported energy-dependent 
X-ray morphologies of SNR 1987A; i.e., the soft X-ray images (0.3 $-$ 
1.2 keV) were correlated with the optical images and the hard band image 
(1.2 $-$ 8.0 keV) was consistent with the radio image. Figure~2 shows 
the development of X-ray spots in both soft and hard bands for the past
17-month period. These images confirm the previous findings that the soft 
X-ray spots are generally coincident with the optical spots while the hard 
X-ray spots are consistent with the radio bright lobes. These correlations 
of X-ray features with optical and radio images are consistent with the 
interpretation that the X-ray emission is from slow and fast shocks 
propagating in the HII region toward the dense inner ring.

The combined line profile as detected with the ACIS/HETG has indicated
a shock front velocity of $\sim$3400 km s$^{-1}$ (Michael et al., 2002),
which was consistent with the SNR expansion rate ($\sim$3500 km s$^{-1}$)
as measured with the radio data (Gaensler et al., 2000) and with the
blast wave velocity inferred by hydrodynamic models (Borkowski et 
al., 1997). We can also estimate the SNR expansion rate by directly
measuring the average radius of the X-ray emission as a function of time.
We fit the radial count distribution with a Gaussian and use the mean
radius of the best-fit Gaussian as the radial extent of the X-ray shell.
Fitting these data to a linear rate, we obtain an X-ray expansion rate
of 4954$\pm$1057 km s$^{-1}$ (Figure~3).

\clearpage

\vspace{-4ex}
\hspace{1.5cm}\includegraphics[angle=-0,width=0.84\textwidth]{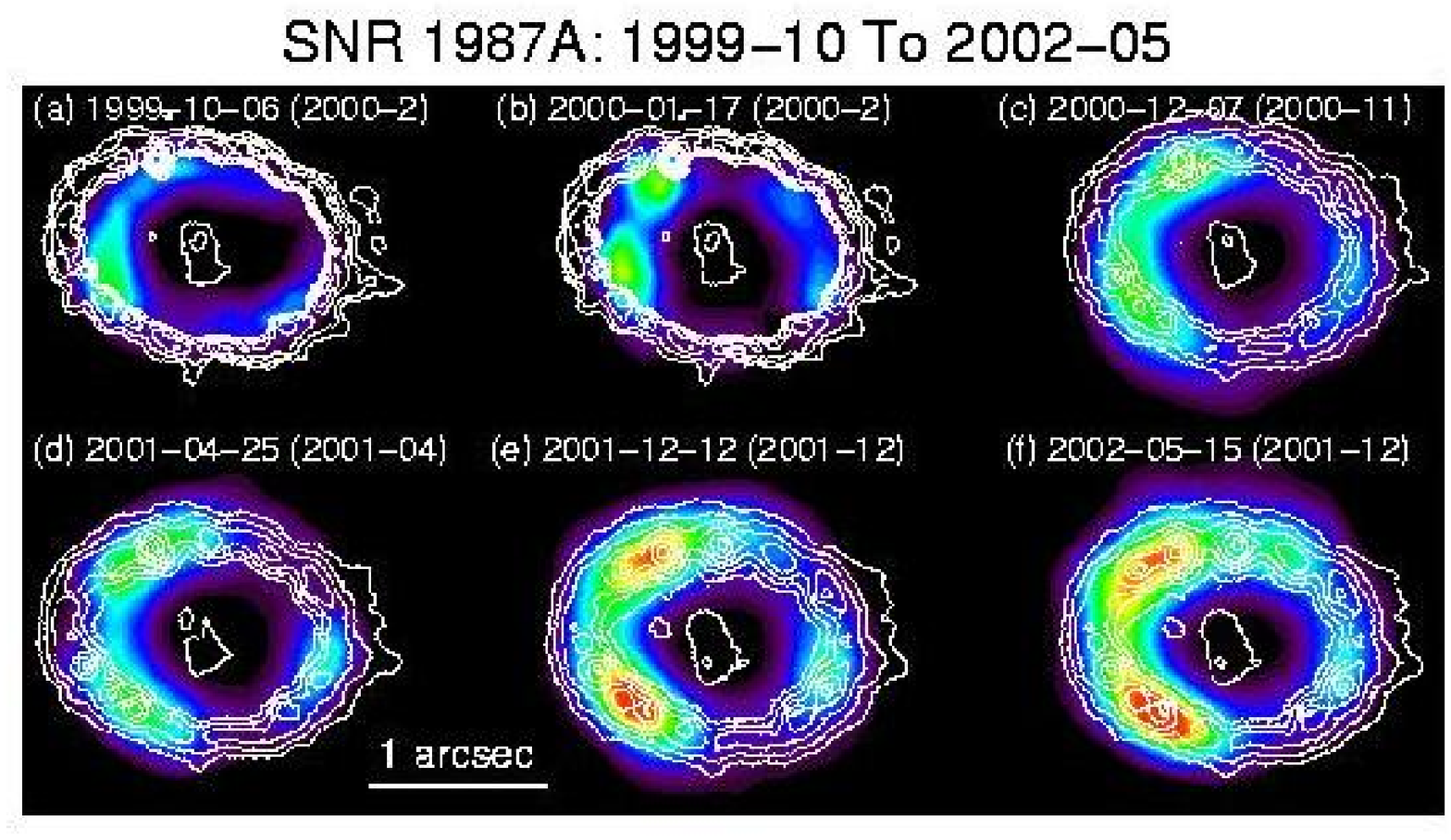}\\
{\sf Fig. 1. ~~Exposure-corrected deconvolved X-ray images of SNR 1987A
overlaid with the HST contours.}
\vspace{3ex}

\hspace{0.8cm}\includegraphics[angle=-0,width=0.85\textwidth]{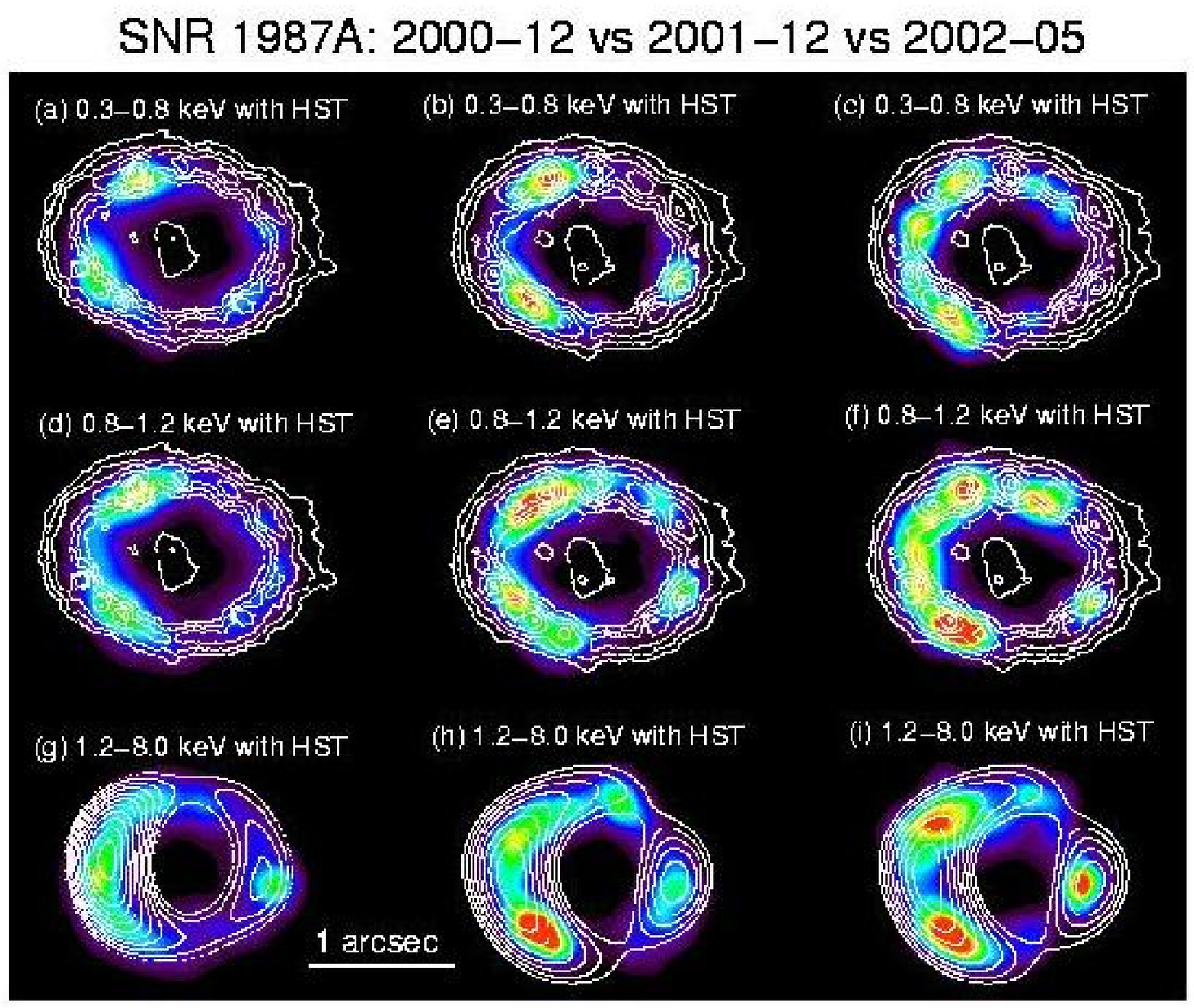}\\
{\sf Fig. 2. ~~Subband images of SNR 1987A. }

\clearpage

\includegraphics[width=\textwidth]{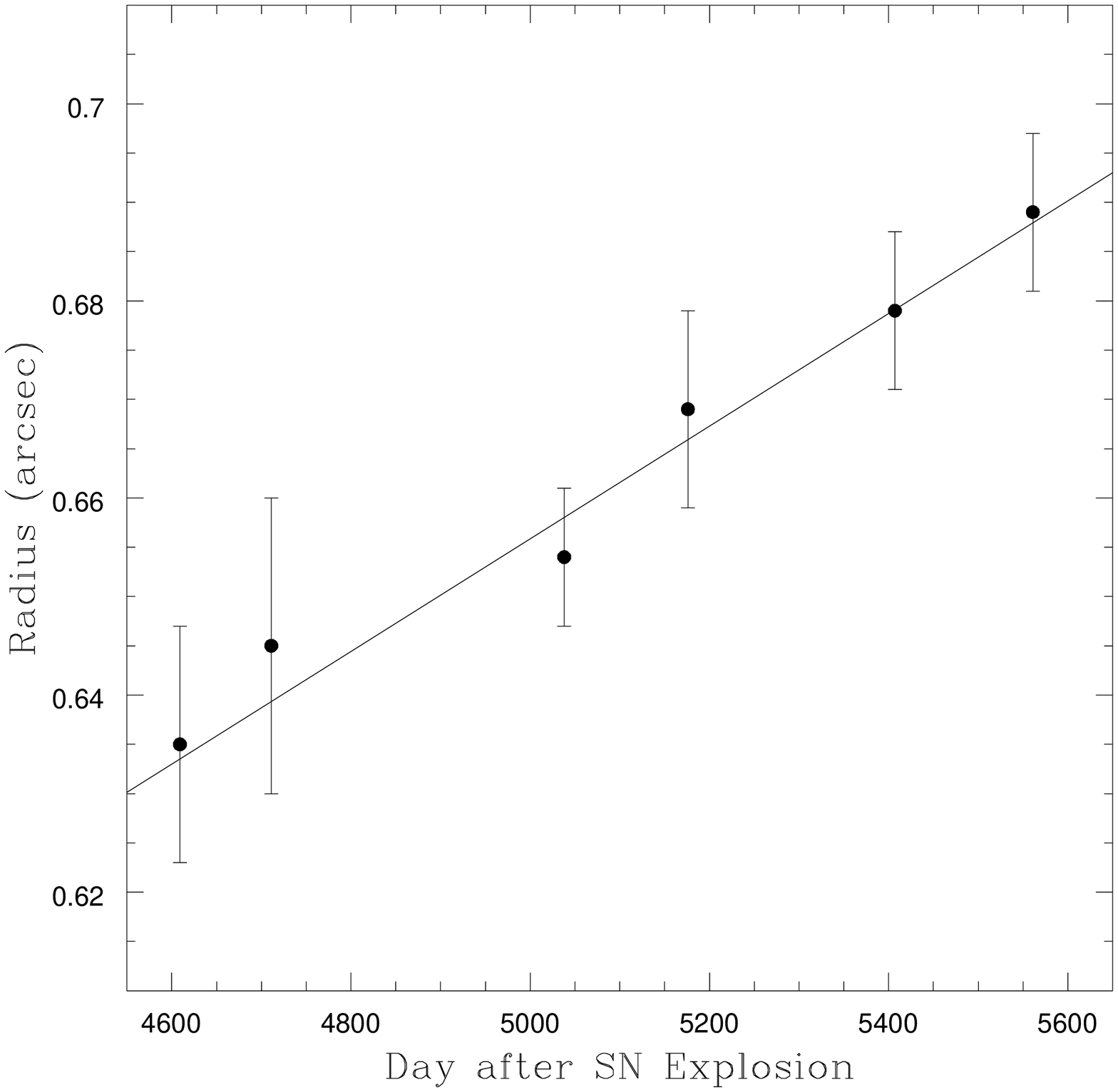}
{\sf Fig. 3. ~~Long-term variation of the mean radius of the X-ray
count distribution. The solid line is the best-fit linear rate
(4954$\pm$1057 km s$^{-1}$).
}
\vspace{1ex}

\section*{SPECTRUM AND LIGHTCURVE}

The undispersed X-ray spectrum from SNR 1987A, as extracted from the
latest observation taken on May 2002, is presented in Figure~4. The
spectra from five other observations show similar features to those in
Figure~4. The X-ray spectrum is thermal and shows broad emission line
features for the elemental species O, Ne, Mg, Si, and S.
The observed spectrum is fitted with a plane-parallel shock with an
electron temperature of $kT$ $\sim$ 2.5 keV and an ionization timescale
of $n_et$ $\sim$ 5 $\times$ 10$^{10}$ cm$^{-3}$ s. The fitted metal
abundances are typically subsolar, which is consistent with the LMC 
ISM (Russell and Dopita, 1992) and ring abundances (Lundqvist and Fransson, 
1996). The fitted electron temperature is consistent within uncertainties 
for all six observations. The volume emission measure ($EM$) has 
constantly increased over the past $\sim$30 months as the SNR has 
brightened. In the bottom panels of Figure~5, we present the $EM$ (left)
and the derived soft/hard X-ray fluxes (right) for the overall spectrum 
between January 2000 and May 2002. We have also separately performed 
the spectral analysis between eastern and western halves and present the 
best-fit $EM$ for each half in the upper panels of Figure~5 (because of 
the limited photon statistics, only the latest four observations were 
feasible for this spectral analysis by halves).  While the overall $EM$ 
has been steadily increasing, the latest data begin to show different 
$EM$ evolutions between the east and the west. This is an interesting 
observation and may indicate some complex density variations of the CSM 
along the inner ring. Follow-up observations will be necessary to verify 
this differential developments of $EM$ between the east and the west. We 
derive the 0.5 $-$ 2.0 keV flux from each observation (Table~2) in order 
to monitor the long-term variations of the soft X-ray flux. The latest 
lightcurves are presented in Figure~6. The X-ray fluxes in Table~2 have 
been corrected for the ACIS quantum efficiency (QE) degradation. For 
comparison, the derived soft X-ray fluxes before (triangles) and after 
(squares) the QE correction are presented in Figure~6b. We note that the 
effects of the QE degradation appear to become significant since the 2000 
December observation. In Figure~6a, we present the radio flux variation, 
which shows a deviation from the linear increase rate since 1997 (around 
Day 4000). The solid lines in Figure~6b represent the linear rates 
fitted to the first four ACIS data points as presented in Park et al. 
(2002). It is evident that the ACIS rate increase has turned up and that 
the ACIS data can no longer be fitted with a linear rate. A quadratic 
increase rate for the combined ({\it ROSAT} + ACIS) lightcurve cannot 
fit the data (the dotted curve in Figure~6b). The X-ray flux has nearly 
tripled for the last $\sim$30 months.

\includegraphics[angle=-0,width=\textwidth]{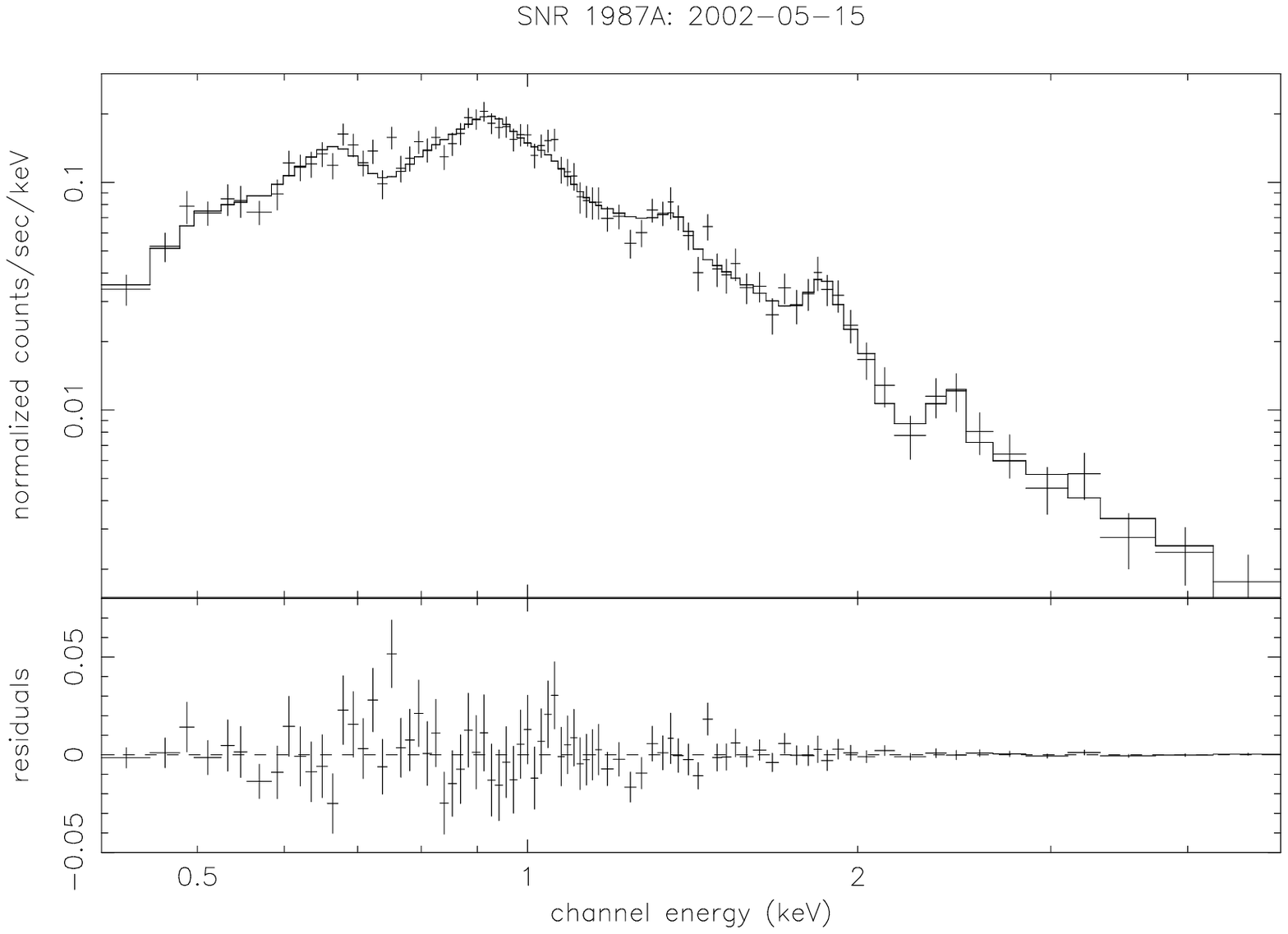}
{\sf Fig. 4. ~~X-ray spectrum of SNR 1987A as observed on 2002-05-15.
}

\vspace{5ex}

\section*{POINT SOURCE}

As of 2002 May, we still observe no direct evidence of a point source
within SNR 1987A. With the photon statistics of the latest data, a
90\% upper limit of the embedded point source counts is 8\% of the total
SNR counts in the 2 $-$ 8 keV band. This implies an {\it observed} upper
limit of $\sim$5.5 $\times$ 10$^{33}$ ergs s$^{-1}$ in the 2 $-$ 10 keV band.

\includegraphics[angle=0,width=\textwidth]{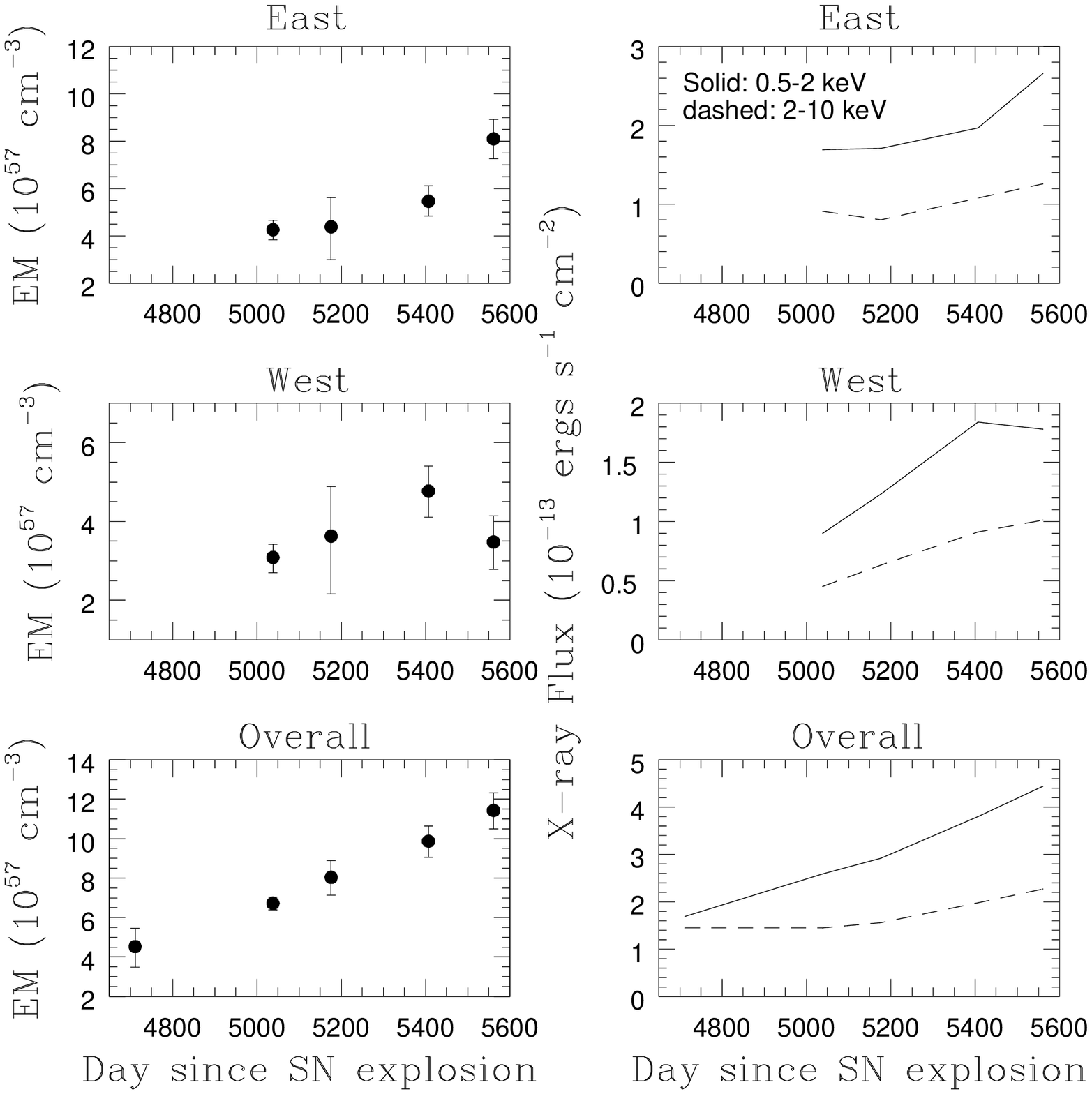}
{\sf Fig. 5. ~~Emission measure/flux variations of SNR 1987A between
2000-01 and 2002-05.
}

\begin{table}[h]
\vspace{-5mm}
\begin{minipage}{90mm}
 \caption{ ~The 0.5 $-$ 2.0 keV Flux and Luminosity of SNR 1987A
from Chandra/ACIS.}
\begin{center}
\begin{tabular}{|ccc|}
\hline
Day$^{a}$ & Observed Flux & Luminosity \\
 & (10$^{-13}$ ergs s$^{-1}$ cm$^{-2}$) & (10$^{35}$ ergs s$^{-1}$) \\
\hline
4609 & 1.62$\pm$0.06 & 0.9\\
4711 & 1.74$\pm$0.07 & 0.9\\
5038 & 2.59$\pm$0.03 & 1.3\\
5176 & 2.93$\pm$0.07 & 1.5\\
5407 & 3.82$\pm$0.05 & 2.0\\
5561 & 4.45$\pm$0.06 & 2.2\\
\hline
\end{tabular}
\end{center}
$^{a}$ Day since the SN explosion.\\ 
\end{minipage}
\hfil\hspace{\fill}
\begin{minipage}{75mm}

\section*{SUMMARY}

Since 1999 October, we have been monitoring the X-ray remnant of SN 1987A
with the high resolution {\it Chandra}/ACIS instrument. With the {\it 
Chandra} observations, we have resolved the shell-like morphology of SNR 
1987A.  For the last 30 months, SNR 1987A has brightened in X-rays by 
nearly a factor of 3. The intensity increase rate has turned up and we 
can no longer fit the long-term lightcurve with a linear rate.  This 
is good evidence of the blast wave entering the main body of the dense 
inner ring. With the total
%\section*{POINT SOURCE}
%
%As of 2002 May, we still observe no direct evidence of a point source
%within SNR 1987A. With the photon statistics of the latest data, a
%90\% upper limit of the embedded point source counts is 8\% of the total
%SNR counts in the 2 $-$ 8 keV band. This implies an {\it observed} upper
%limit of $\sim$5.5 $\times$ 10$^{33}$ ergs s$^{-1}$ in the 2 $-$ 10 keV band.
\end{minipage}
\end{table}

\includegraphics[angle=0,width=\textwidth]{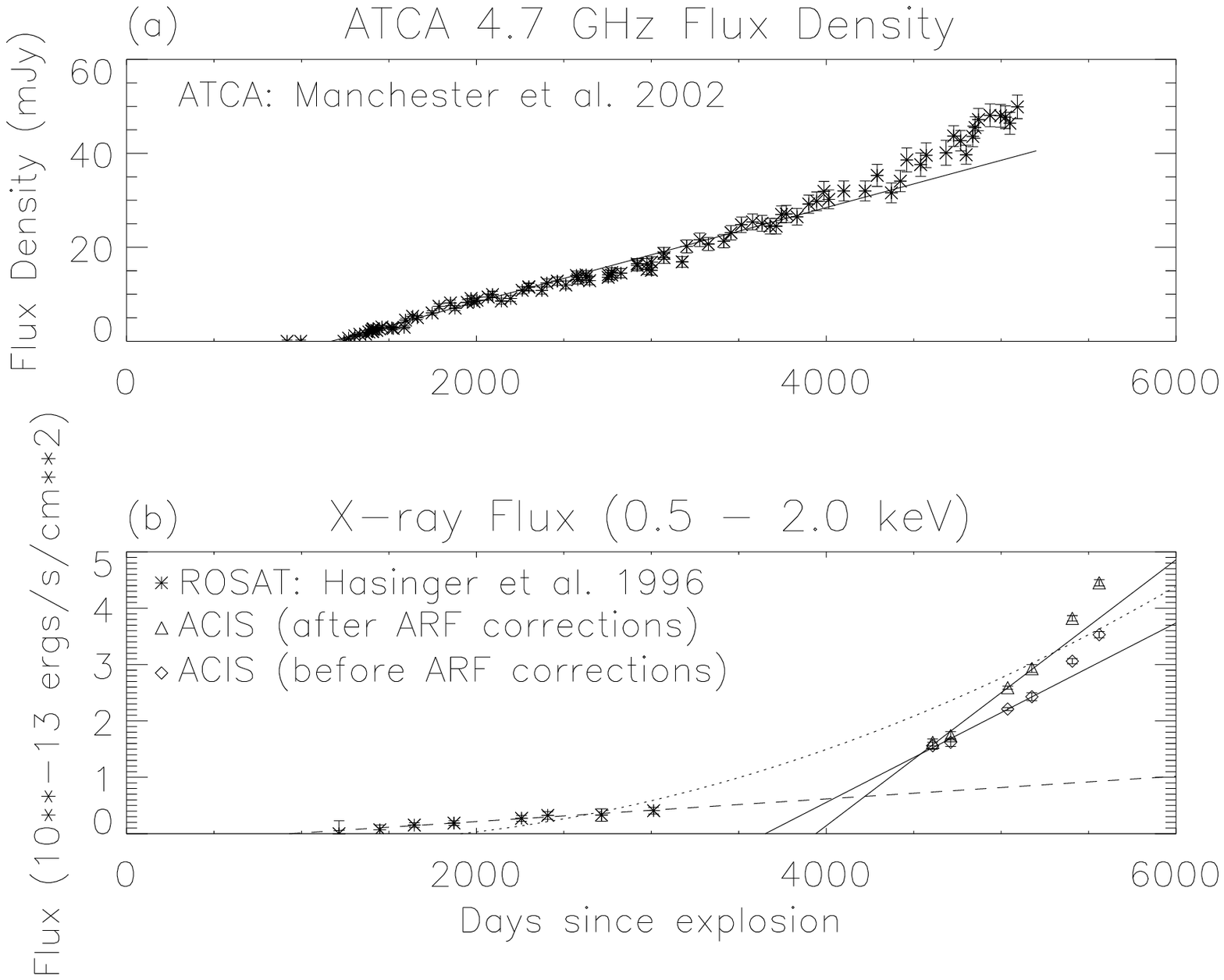}
{\sf Fig. 6. ~(a) The long-term lightcurve of SNR 1987A in the 4.7 GHz
(ATCA). (b) The long-term X-ray lightcurve.
}

\vspace{5mm}

%\section*{SUMMARY}
%
%Since 1999 October, we have been monitoring the X-ray remnant of SN 1987A
%with the high resolution {\it Chandra}/ACIS. With the {\it Chandra}
%observations, we have resolved the shell-like morphology of SNR 1987A.
%For the last 30 months, SNR 1987A has brightened in X-rays by nearly a 
%factor of 3. The intensity increase rate has turned up and we can no longer 
%fit the long-term lightcurve with a linear rate.  This is good evidence 
%of the blast wave entering the main body of the dense inner ring. With 
\noindent of six observations now, we detect newly developing X-ray spots. 
The soft X-ray spots are consistent with the optical spots while the hard 
X-ray spots are coincident with the radio lobes. These characteristics 
of the X-ray spots support the interpretation that the soft spots are the 
X-ray emission from the slow shock encountering dense protrusions of the 
inner ring and that the hard spots are emission from a fast forward shock 
propagating through the HII region. The emergence of new X-ray 
spots in the western side in addition to the eastern side indicates 
that the X-ray emission from SNR 1987A will be a {\it complete} ring 
in near future. With two new ACIS observations, we confirm the 
previously reported radial expansion rate of the X-ray SNR. The spectrum 
is fitted with a plane-parallel shock model with $kT$ $\sim$ 2.5 keV and 
low abundances, which is consistent with the previous results indicating 
X-ray emission primarily from the shocked CSM. The latest observation 
suggests a differential evolution of the volume emission measure between 
the east and the west, which might be an indication of complex density 
structure of the CSM along the inner ring. The current data are however
insufficient to make a firm conclusion and follow-up observations should
be helpful to understand this feature. We obtain a 90\% upper limit on the
2 $-$ 10 keV band X-ray luminosity $L_X$ = 5.5 $\times$ 10$^{33}$ ergs 
s$^{-1}$ for any embedded point source.

\section*{ACKNOWLEDGEMENTS}

The authors thank P. Challis and the SINS collaboration for providing
the {\it HST} data. We also thank B. Gaensler and the Australian
Telescope Compact Array for providing the radio data.
This work has been supported by NASA under the contract NAG8-01128
and SAO grants GO1-2064B and GO2-3098A.

\bibliographystyle{natbib}

%\clearpage
%
%\vspace{-4ex}
%\hspace{1.5cm}\includegraphics[angle=-0,width=0.84\textwidth]{fig1.ps}
%
%{\sf Fig. 1. ~~Exposure-corrected false-color images of SNR 1987A
%overlaid with the HST contours.}
%\vspace{3ex}
%
%\hspace{0.8cm}\includegraphics[angle=-0,width=0.85\textwidth]{fig2.ps}
%
%{\sf Fig. 2. ~~Subband images of SNR 1987A. }

\end{document}